\documentclass[journal]{IEEEtranTCOM}
\normalsize

\usepackage[T1]{fontenc}% optional T1 font encoding
\ifCLASSINFOpdf
\else
\fi
\usepackage{amsmath,amssymb}
\usepackage{graphicx}
\usepackage{amsthm}
\theoremstyle{definition}

\newtheorem*{theorem*}{Theorem}

\usepackage{txfonts}
\usepackage{setspace}
\begin{document}
%\title{Non-Linear Programming Maximizing SNR\\for  Designing Spreading Sequence}
%\title{A SNR Maximization as a Non-Linear Programming}
\title{Non-Linear Programming: Maximize SNR\\ for Designing Spreading Sequence -- Part I:\\ SNR versus Mean-Square Correlation}
%An Associated Expression of SNR}

\author{Hirofumi~Tsuda ~\IEEEmembership{Student Member,~IEEE,} Ken Umeno
%~\IEEEmembership{Fellow,~OSA}
        %and~Jane~Doe,~\IEEEmembership{Life~Fellow,~IEEE}% <-this % stops a space
\thanks{H. Tsuda and K. Umeno are with the Department
of Applied Mathematics and Physics, Graduate School of Informatics, Kyoto University, Kyoto, 606-8561 Japan (email: tsuda.hirofumi.38u@st.kyoto-u.ac.jp, umeno.ken.8z@kyoto-u.ac.jp).}}% <-this % stops a space


%Telephone: (800) 555--1212, Fax: (888) 555--1212}

\markboth{IEEE Transactions on Communications}%
{Submitted paper}


\maketitle

\begin{abstract}
Signal to Noise Ratio (SNR) is an important index for wireless communications. In CDMA systems, spreading sequences are utilized. This series of papers show the method to derive spreading sequences as the solutions of the non-linear programming: maximize SNR. In this paper, we consider a frequency-selective wide-sense-stationary uncorrelated-scattering (WSSUS) channel and evaluate the worst case of SNR. Then, we derive the new expression of SNR whose main term consists of the periodic correlation terms and the aperiodic correlation terms. In general, there is a relation between SNR and mean-square correlations, which are indices for performance of spreading sequences. Then, we show the relation between our expression and them. With this expression, we can maximize SNR with the Lagrange multiplier method. In Part II, with this expression, we construct two types optimization problems and evaluate them. 
\end{abstract}

\begin{IEEEkeywords}
Asynchronous CDMA, Spreading sequence, Rician fading, Signal to noise ratio, Non-Linear Programing
\end{IEEEkeywords}

% \fi
\IEEEpeerreviewmaketitle

\section{Introduction}
\IEEEPARstart{S}{preading sequences} are utilized in code division multiple access (CDMA) systems, which is one of the Multiple access systems \cite{multiple}. The one of CDMA systems, Direct Sequence CDMA (DS-CDMA) \cite{dscdma} is used for the 3G mobile communication system. In CDMA systems, we use spreading sequences to modulate and demodulate signals. Therefore, spreading sequences are necessary to communicate in CDMA systems.

To improve CDMA systems, there are many works of designing spreading sequences. The aim of designing spreading sequences is to make Signal to Noise Ratio (SNR) high. It is necessary and sufficient for achieving the spectral efficiency to increase the Signal to Noise Ratio (SNR) \cite{efficiency}. The current spreading sequences are the Gold codes \cite{gold}. These sequences are obtained from M-sequences. Therefore, the Gold codes are obtained from shift registers. In \cite{chaos_cdma}-\cite{mazzini}, it is proposed to use chaotic dynamical systems to design spreading sequences. For these chaos-based DS-CDMA systems, the performance in fading channels is investigated in \cite{kaddoum}-\cite{indoor}. Their approaches to obtain spreading sequences are to design the system which generates sequences.

Other approaches are to derive sequences which satisfy the equality of the limitation. In CDMA systems, crosscorrelation is treated as a basic component of interference noise and autocorrelation is related to synchronization at the receiver side and the fading noise, thus, it is desirable that the first peak of crosscorrelation and the second peak in autocorrelation should be kept low. Sarwate \cite{sarwate} has shown that there is an avoidable limitation trade-off as a relation between lowering crosscorrelation peak and autocorrelation's second peak. The FZC sequences \cite{zadoff} \cite{chu} satisfy the equality of the limitation. Welch \cite{welch} shows that the maximum value of crosscorrelation is bounded below. This limitation is called the Welch bound and the sequences which satisfy the equality of the Welch bound are called as the Welch Bound Equality (WBE) sequences. The WBE sequences have been investigated in \cite{wbe} \cite{general_wbe}.

In contrast, our approach is to derive directly sequences whose SNR is high. We consider a Rician fading channel, and evaluate the worst case of SNR and derive spreading sequences as solutions of the optimization problem: maximize SNR. Therefore, our spreading sequences are guaranteed to have high SNR. The expression of SNR has been obtained in \cite{mazzini} and \cite{pursley}. However, their expressions are not differentiable since they have the real part operator. Therefore, it is not straightforward to solve the optimal problem with their expressions, and then a differentiable expression of SNR has been demanded. In this paper, we derive the differentiable expression of SNR, which does not have the real part operator. Moreover, the main term of our expression consists of the periodic correlation terms and the aperiodic correlation terms. This result shows that there is the clear relation among SNR, the periodic and aperiodic correlation. In Part II, to have a such a expression, we consider two types of problems: maximize the average of SNR and maximize the minimum SNR. With our expression, we can numerically solve the problems and obtain the solutions. 

This paper is organized as follows. In Section II, we show an asynchronous CDMA system model. In this model, we assume a Rician fading channel. This model is general and has been studied in \cite{pursley} \cite{borth} and \cite{mazzini}. Then, in Section III, we make some assumptions and evaluate SNR which is worst case. This situation is equivalent to that the effect of multipath fading is the largest. In Section VI, we derive the new expression of SNR. To derive it, we use the two types of orthogonal basis vectors in correlation. In section V, we evaluate our expression of SNR. In general, it is necessary to reduce mean-square correlations for high SNR \cite{fzcfam} \cite{meansquare}. This section shows the relation between our expression and mean-square correlations. Finally,  some conclusions are drawn and directions of further investigations are discussed.

\section{Asynchronous CDMA Model}
In this section, we fix our model used thorough this paper and mathematical symbols that will be used in the following sections. We consider the following asynchronous binary phase shift keying (BPSK) CDMA model \cite{pursley} \cite{borth}. Let $N$ be the length of spreading sequences. The user $k$'s data signal $b_k(t)$ is expressed as
\begin{equation}
b_k(t) = \sum_{n=-\infty}^{\infty} b_{k,n} p_{T}(t - nT),
\end{equation}
where $b_{k,n} \in \{-1,1\}$ is the $n$-th component of bits which the user $k$ send, $T$ is the duration of one symbol and $p_{T}(t)$ is a rectangular pulse written as
\[p_T(t) = \left\{ \begin{array}{c c}
1 & 0 \leq t \leq T\\
0 & \mbox{otherwise}
\end{array} \right. .\]
The user $k$'s code waveform $s_k(t)$ is expressed as
\begin{equation}
s_k(t) = \sum_{n=-\infty}^{\infty} s_{k,n} p_{T_c}(t - nT_c),
\end{equation}
where $s_{k,n}$ is the $n$-th component of the user $k$'s spreading sequence and $T_c$ is the width of the each chip such that $NT_c = T$. We assume that the sequence $(s_{k,n})$ has the period $N$, that is, $s_{k,n}=s_{k,n+N}$. Moreover, we assume the condition that
\begin{equation}
\sum_{n=1}^N \left|s_{k,n}\right|^2 = N.
\end{equation}
This is proven in the appendix A. This condition is often used \cite{sarwate} \cite{welch}.

The user $k$'s transmitted signal $\zeta_k(t)$ is
\begin{equation}
\zeta_k(t) = \sqrt{2P} \operatorname{Re}[s_k(t)b_k(t)\exp(j \omega_c t + j\theta_k)],
\label{eq:carrer}
\end{equation}
where $P$ is the common signal power, $\omega_c$ is the common carrier frequency and $\theta_k$ is the phase of the user $k$.

We consider a Rician fading channel. The received signal $r(t)$ is
\begin{equation}
r(t) = \sum_{k=1}^K \operatorname{Re}\left[u_k(t - \tau_k)\exp\left(j \omega_c t + j \psi_k \right) \right] + n(t),
\end{equation}
where $\psi_k = \theta_k - \omega_c \tau_k$, $n(t)$ is the additive white Gaussian noise (AWGN) and $u_k$ is
\begin{equation}
u_k(t) = \gamma_k \int_{-\infty}^{\infty} h_k(\tau, t)x_k(t-\tau)d\tau + x_k(t),
\label{eq:rice}
\end{equation}
\begin{equation}
x_k(t) = \sqrt{2P} s_k(t)b_k(t).
\label{eq:rice2}
\end{equation}
The first term of Eq. (\ref{eq:rice}) is the component of faded signals and the second term is the component of a direct wave. The function $h_k(\tau, t)$ is the zero-mean complex Gaussian random process and $\gamma_k$ is the nonnegative real parameter which represents the transmission coefficient for the user $k$'s signal. In general, $h(\tau, t)$ is often approximated by \cite{ofdmandcdma} \cite{peter} \cite{variance}
\begin{equation}
h_k(\tau, t) = \frac{1}{\sqrt{N}}\sum_{n=1}^{N} a^{(k)}_n \exp\left(2 \pi j f_n(t) \right)\delta(\tau - \tau^{(k)}_n),
\label{eq:h_app}
\end{equation}
where
\[f_n(t) = f^{(k)}_{D_n} t + f^{(k)}_{\operatorname{hop}}(t)\tau^{(k)}_n,\]
$a^{(k)}_n$ is the attenuation coefficient, $f^{(k)}_{D_n}$ is the Doppler frequency and $f^{(k)}_{\operatorname{hop}}(t)$ is the carrier frequency shift, $\tau^{(k)}_n$ is the delay time of the $n$-th delayed signal and $\delta(x)$ is the delta function. 

If the received signal $r(t)$ is the input to a correlation receiver matched to $\zeta_i(t)$, then the corresponding output $Z_i$ is
\begin{equation}
Z_i = \int_{0}^T r(t) \operatorname{Re}[s_i(t-\tau_i)\exp(j \omega_c t + j \psi_i)]dt.
\label{eq:output}
\end{equation}
Without loss of generality, we assume $\tau_i = 0$ and $\theta_i = 0$ and hence $\psi_i = 0$. With a low-pass filter, we can ignore double frequency terms, and rewrite Eq. (\ref{eq:output}) as
\begin{equation}
\begin{split}
Z_i &= \frac{1}{2} \sum_{k=1}^K \int_{0}^T \operatorname{Re}[u_k(t- \tau_k)\overline{s_i(t)}\exp(j \psi_k)]dt\\
&+ \int_0^T n(t)\operatorname{Re}[s_i(t)\exp(j \omega_c t)],
\label{eq:output2}
\end{split}
\end{equation}
where $\overline{z}$ is complex conjugate of $z$ and 
\begin{equation}
\overline{s_i(t)} = \sum_{n=-\infty}^{\infty} \overline{s_{i,n}} p_{T_c}(t - nT_c).
\end{equation}
In obtaining  Eq. (\ref{eq:output2}), we have used the identity
\begin{equation}
2\operatorname{Re}[z_1]\operatorname{Re}[z_2] = \operatorname{Re}[z_1z_2] + \operatorname{Re}[z_1\overline{z_2}],
\label{eq:thm1}
\end{equation}
where $z_1, z_2 \in \mathbb{C}$.

Similar to \cite{pursley}, we assume that the phase $\psi_k$, time delays $\tau_k$ and symbols $b_{k,n}$ are independent random variables and they are uniformly distributed on $[0, 2\pi)$, $[0,T)$ and $\{-1 ,1\}$. Without loss of generality, we assume that $b_{i,0} = +1$. 

To evaluate SNR, we define
\begin{equation}
\mu_{i,k}(\tau; t) = b_k(t - \tau)s_k(t - \tau)\overline{s_i(t)}
\end{equation}
and
\begin{equation}
\xi_{i, k_1,k_2}(\tau_1, \tau_2; t_1, t_2) = \mu_{i, k_1}(\tau_1; t_1) \overline{\mu_{i, k_2}}(\tau_2; t_2).
\end{equation}
For notational convenience, we write $\mu_{i,i}$ as $\mu_i$, and $\xi_{i,i,i}$ as $\xi_i$. We divide $Z_i$ into the four signals, the user $i$'s desired signal $D_i$, the user $i$'s faded signal $F_i$, the interference signal $I_i$ and the AWGN signal $N_i$. They are expressed as
\begin{equation}
\begin{split}
D_i &= \sqrt{\frac{P}{2}} \int_0^T b_i(t)dt\\
F_i &= \sqrt{\frac{P}{2}}\operatorname{Re}[\tilde{F}_i]\\
I_i &= \sqrt{\frac{P}{2}} \sum_{\substack{k=1 \\ k \neq i}}(\operatorname{Re}[\gamma_k \tilde{I}_{i,k}] + \operatorname{Re}[\tilde{I}'_{i,k}])\\
N_i &= \int_0^Tn(t)\operatorname{Re}[s_i(t)\exp(j\omega_ct)]
\end{split}
\end{equation}
where
\begin{equation*}
\begin{split}
\tilde{F}_i & = \int_0^T \int_{-\infty}^{\infty}\gamma_i h_i(\tau,t) \mu_i(\tau ; t) d\tau dt\\
\tilde{I}_{i,k} &= \int_0^T \int_{-\infty}^{\infty} h_k(\tau,t - \tau_k) \mu_{i,k}(\tau_k + \tau ; t) \exp(j \psi_k) d \tau d t\\
\tilde{I}'_{i,k} &= \int_0^T \mu_{i,k}(\tau_k; t) \exp(j \psi_k) dt.\\
\end{split}
\end{equation*}
From these expressions, $Z_i$ is expressed as
\begin{equation}
Z_i = D_i + F_i + I_i + N_i.
\end{equation}
\section{Evaluation of SNR}
Since $\operatorname{E}\{F_i\} = \operatorname{E}\{I_i\} = \operatorname{E}\{N_i\} = 0$ and $\displaystyle\operatorname{E}\{D_i\} = T\sqrt{P/2}$, we have $\displaystyle \operatorname{E}\{Z_i\} = T\sqrt{P/2}$, where $\operatorname{E}\{X\}$ is the average of $X$. We assume that the each Gaussian process $h_k(\tau, t)$ is independent and $h_k(\tau, t)$, $\psi_k$, $\tau_k$ and $b_{k,n}$ are independent. Then, SNR of the user $i$ is defined as
\begin{equation}
\operatorname{SNR}_i =\sqrt{\frac{\operatorname{Var}\{D_i\} }{\operatorname{Var}\{F_i\}  + \operatorname{Var}\{I_i\}  + \operatorname{Var}\{N_i\} }}.
\label{eq:SNR_def}
\end{equation}
In this section, we focus on the estimation of the lower bound of Eq. (\ref{eq:SNR_def}) under some assumptions.
It is known from \cite{pursley} and \cite{borth} that the variance of $N_i$ is
\begin{equation}
\operatorname{Var}\{N_i\} = \frac{1}{4}N_0T
\end{equation}
if $n(t)$ has a two-sided spectral density denoted as $\frac{1}{2}N_0$.

We make assumptions about the channel, the variables and the Gaussian process $h_k(\tau,t)$ that
\begin{enumerate}
\item the Fourier transform of $h_k(\tau, t)$ and its inverse Fourier transform exist. 
\item the channel is a wide-sense-stationary uncorrelated-scattering (WSSUS) channel \cite{bello}.
\item the channel is a frequency selective fading channel.
\item the Gaussian process $h_k(\tau,t)$ satisfies $h_k(\tau,t)=0$ when $\tau < 0$ \cite{variance} .
\item there is an integer $M_k$ that satisfies $h_k(\tau,t)=0$ when $\tau > M_kT$.
\item the variable $\tau$ satisfies that $n T + l T_c \leq \tau_k < nT + (l + 1)T_c$, where $n$ and $l$ are the integers which satisfy $0 \leq n$ and $0 \leq l < N$.
\item the phase $\psi_k$, time delays $\tau_k$ and symbols $b_{k,n}$ are independent random variables and they are uniformly distributed on $[0, 2\pi)$, $[0,T)$ and $\{-1 ,1\}$, respectively.
\end{enumerate}
The first assumption is required to define a WSSUS channel. The second and third assumptions are often used in the analysis of wireless communications. The fourth assumption is equivalent to that the channel is causal. The fifth assumption is equivalent to the one that the delayed signal becomes to zero in finite-time. From Eq. (\ref{eq:h_app}), the faded signal is composed of the sum of the delayed signal which is affected by the Doppler shift and the delayed signals. They are attenuated as time passes. The models that the probability of time delay $\tau$ obeys an exponential distribution are often used \cite{peter}. The sixth assumption is often used \cite{pursley} \cite{borth}. The last assumption is written in Section II.

In WSSUS channels, the covariance function of $h_k(\tau, t)$ is expressed as \cite{borth}
\begin{equation}
\begin{split}
\Sigma_k(\tau_1, \tau_2; t_1, t_2) &= \operatorname{E}[h_k(\tau_1,t_1)\overline{h_k(\tau_2,t_2)}]\\
&=\rho_k(\tau_1, t_1-t_2)\delta(\tau_1 - \tau_2).
\end{split}
\end{equation}
Adding to this condition, in a selective fading channel, covariance function $\Sigma_k$ is \cite{borth}
\begin{equation}
\begin{split}
\Sigma_k(\tau_1, \tau_2; t_1, t_2) &= \rho_k(\tau_1, 0)\delta(\tau_1 - \tau_2)\\
&= g_k(\tau_1)\delta(\tau_1 - \tau_2).
\label{eq:covariance}
\end{split}
\end{equation}
In the above equation, we have defined $g_k(\tau_1) = \rho_k(\tau_1, 0)$. From Eq. (\ref{eq:covariance}), the covariance function $\Sigma_k$ is independent of $t_1$ and $t_2$.

First, we calculate the variance of $F_i$. With Eq. (\ref{eq:thm1}), $\operatorname{Var}\{F_i\}$ is
\begin{equation}
\begin{split}
\operatorname{Var}\{F_i\} &= \frac{P}{2}\operatorname{E}\{\operatorname{Re}[\tilde{F_i}]^2\}\\
&=\frac{P}{4}\operatorname{E}\{\operatorname{Re}[\tilde{F_i}^2]\} + \frac{P}{4}\operatorname{E}\{|\tilde{F_i}|^2\}.
\end{split}
\end{equation}
Here, $\operatorname{E}\{\operatorname{Re}[\tilde{F_i}^2]\}$ and $\operatorname{E}\{|\tilde{F_i}|^2\}$ are expressed as
\begin{equation}
\begin{split}
\operatorname{E}\{\operatorname{Re}[\tilde{F_i}^2]\} =& \gamma_i^2 \cdot \operatorname{E}_{\mathbf{b}_i}\left\{ \operatorname{Re}\left[\int_0^T\int_0^T \int_{-\infty}^\infty \int_{-\infty}^\infty\tilde{\Sigma}_i(\tau_1,\tau_2;t_1,t_2)\right. \right. \\
&\left. \cdot \mu_{i}(\tau_1; t_1)\mu_{i}(\tau_2; t_2) d\tau_1 d\tau_2 dt_1 dt_2 \Bigg]\right\},\\
\operatorname{E}\{|\tilde{F_i}|^2\} =& \gamma_i^2 \cdot \operatorname{E}_{\mathbf{b}_i}\Bigg\{ \int_0^T\int_0^T \int_{-\infty}^\infty \int_{-\infty}^\infty\Sigma_i(\tau_1,\tau_2;t_1,t_2)\\
&\cdot \xi_i(\tau_1,\tau_2;t_1,t_2)d\tau_1 d\tau_2 dt_1 dt_2 \Bigg\},
\end{split}
\end{equation}
where 
\[\tilde{\Sigma}_i(\tau_1,\tau_2;t_1,t_2) = \operatorname{E}\left\{h_i(\tau_1,t_1)h_i(\tau_2,t_2)\right\}\]
and $\operatorname{E}_{\mathbf{b}_i}\{X\}$ is the average over all the bits of the user $i$. We write the variable over which we take the average at the right bottom of $\operatorname{E}$.
In \cite{borth} and \cite{ofdmandcdma}, it is shown that we can use
\begin{equation}
\operatorname{E}\{h_k(\tau_1,t_1)h_k(\tau_2,t_2)\} = 0.
\label{eq:rf}
\end{equation}
This result is obtained from the demodulation of RF signals.
From Eqs. (\ref{eq:covariance})-(\ref{eq:rf}), we have
\begin{equation}
\begin{split}
\operatorname{Var}\{F_i\} =& \frac{P}{4}\gamma_i^2  \cdot \operatorname{E}_{\mathbf{b}_i}\Bigg\{ \int_{-\infty}^\infty g_i(\tau)\\
&\cdot \int_0^T\int_0^T \xi_i(\tau, \tau ; t_1, t_2) dt_1dt_2d\tau \Bigg\}.
\end{split}
\label{eq:fad}
\end{equation}
The double integral term is written as
\begin{equation}
\begin{split}
&\int_0^T\int_0^T \xi_i(\tau, \tau ; t_1, t_2)dt_1dt_2\\
 =& \left(\int_0^T \mu_i(\tau; t_1) dt_1\right)\left(\int_0^T \overline{\mu_i(\tau; t_2)} dt_2\right)\\
=& \left|\int_0^T \mu_i(\tau; t) dt\right|^2 = \Gamma_i(\tau) \geq 0,
\end{split}
\end{equation}
where $\Gamma_i(\tau)$ has been defined. Note that $\Gamma_i(\tau)$ is the squared absolute value of the correlation in an asynchronous CDMA system.
From the assumptions 4 and 5, we obtain 
\begin{equation}
\begin{split}
g_i(\tau) &= 0 \hspace{3mm}\mbox{for} \hspace{2mm}\tau < 0,\\
g_i(\tau) &= 0 \hspace{3mm}\mbox{for} \hspace{2mm}\tau > M_iT.
\end{split}
\end{equation}
It is clear that $g_i(\tau)$ is non negative since
\[g_i(\tau) = \rho_i(\tau,0) = \operatorname{E}\{h_i(\tau,0) \overline{h_i(\tau,0)}\}\geq0.\]
Further, we can assume that $g_i(\tau)$ has the upper bound $C_i$ in $[0,M_iT]$. This is proven in the appendix B. We have no knowledge about the form of $g_i(\tau)$. For this reason, we evaluate the upper bound of $\operatorname{Var}\{F_i\}$ with the product of two terms, one is related to $g_i(\tau)$ and the other is related to the spreading sequences. From H\"older's inequality, we evaluate Eq. (\ref{eq:fad}) as
\begin{equation}
\begin{split}
\operatorname{Var}\{F_i\} =& \frac{P}{4}\gamma_i^2 \cdot \operatorname{E}_{\mathbf{b}_i}\left\{ \int_{-\infty}^\infty g_i(\tau) \Gamma_i(\tau) d\tau\right\} \\
\leq& \frac{P}{4}\gamma_i^2 \cdot \operatorname{E}_{\mathbf{b}_i}\left\{ \sup_{[0,M_iT]}\{|g_i(\tau)|\} \cdot \int_{0}^{M_iT}  |\Gamma_i(\tau)| d\tau \right\}\\
=&  \frac{P}{4}\gamma_i^2 C_i \cdot \operatorname{E}_{\mathbf{b}_i}\left\{ \int_{0}^{M_iT}  \Gamma_i(\tau) d\tau \right\}.
\end{split}
\label{eq:var_f}
\end{equation}
The equality is attained if $g_i(\tau)$ is the rectangular function. This is the worst case where $\operatorname{Var}\{F_i\}$ is maximized. From the assumption 6, the time delay $\tau$ satisfies $n_i T + l_i T_c \leq \tau < n_iT + (l_i + 1)T_c$, where $n_i$ and $l_i$ are the integers which satisfy $0 \leq n_i < M_i$ and $0 \leq l_i < N$. Note that $n_i T + NT_c = (n_i + 1)T$. Since the correlation in an asynchronous CDMA system is the superposition of the correlations in a chip-synchronous CDMA system, the function $\Gamma_i(\tau)$ can be written as
 \begin{equation}
\begin{split}
\Gamma_i(\tau) &= \left|\int_0^T \mu_i(\tau; t) dt\right|^2 \\
&= \left|R_i\left(\tau, n_i, l_i\right) + \hat{R}_i\left(\tau, n_i, l_i\right) \right|^2
\end{split}
\end{equation}
where
 \begin{equation}
 \begin{split}
 R_i\left(\tau, n, l \right) =& \left(\tau - nT - lT_c\right)\\
&\cdot \left\{b_{i,-n-1} \sum_{m=1}^{l}\overline{s_{i,m}}s_{i,N-l+m}+ b_{i,-n}\sum_{m=1}^{N-l} \overline{s_{i,l+m}}s_{i,m} \right\}\\
 \hat{R}_i\left(\tau, n, l\right) =& \left(nT + (l+1)T_c - \tau\right),\\
&\cdot \left\{b_{i,-n-1} \sum_{m=1}^{l+1}\overline{s_{i,m}}s_{i,N-l+m-1}\right.\\
&\left.+ b_{i,-n}\sum_{m=1}^{N-l-1} \overline{s_{i,l+m+1}}s_{i,m} \right\}.
 \end{split}
 \label{eq:cor}
 \end{equation}
Note that $R_i(\tau,n,l)$ and $\hat{R}_i(\tau,n,l)$ are expressed as the autocorrelation function in the chip-synchronous CDMA systems.
From Eq. (\ref{eq:cor}), it is sufficient to consider only two adjacent bits, $b_{i, -n_i-1}$ and $b_{i, -n_i}$. From the independence of each bit $b_{i,-n_i}$, Eq. (\ref{eq:var_f}) can be written as
 \begin{equation}
 \begin{split}
 \operatorname{Var}\{F_i\} &\leq \frac{P}{4}\gamma_i^2 C_i \cdot \operatorname{E}_{\mathbf{b}_i}\left\{ \int_{0}^{M_iT}  \Gamma_i(\tau) d\tau\right\} \\
 &= \frac{P}{4}\gamma_i^2 C_i \cdot \operatorname{E}_{\mathbf{b}_i}\left\{ \sum^{M_i-1}_{n_i=0} \sum^{N-1}_{l_i=0}\int_{n_i T + l_i T_c}^{ n_iT + (l_i + 1)T_c} \Gamma_i(\tau,n_i,l_i) d\tau \right\}\\
 &= \frac{P}{4}\gamma_i^2 C_i M_i \cdot \operatorname{E}_{\mathbf{b}_i}\left\{ \sum^{N-1}_{l_i=0}\int_{l_i T_c}^{(l_i + 1)T_c} \Gamma_i(\tau,0,l_i) d\tau\right\},
 \end{split}
 \label{eq:fad_cor}
 \end{equation}
 where
 \begin{equation*}
 \Gamma_i(\tau,n,l) = \left|R_i\left(\tau, n, l\right) + \hat{R}_i\left(\tau, n, l\right) \right|^2.
 \end{equation*}
Since $P\gamma_i^2 C_i M_i$ is a constant, it is sufficient to focus on the sum term in the right hand side of Eq. (\ref{eq:fad_cor}) to reduce the upper bound of $ \operatorname{Var}\{F_i\}$.

Similar to the fading term, we evaluate the interference noise term $I_i$. The variance of $I_i$ is
\begin{equation}
\operatorname{Var}\{I_i\} =\frac{P}{4}\sum^K_{\substack{k=1 \\ k \neq i}}\left[\gamma_k^2 \operatorname{Var}\{|\tilde{I}_{i,k}|\} + \operatorname{Var}\{|\tilde{I}'_{i,k}|\}\right].
\label{eq:decomp}
\end{equation}
 In the above equation, we have used Eq. (\ref{eq:thm1}) and Eq. (\ref{eq:rf}). It is clear that
\begin{equation}
E_{\psi_k}\left\{\operatorname{Re}\left[\left(\tilde{I}'_{i,k}\right)^2\right] \right\} = 0.
\end{equation} 
In Eq. (\ref{eq:decomp}), $\tilde{I}_{i,k}$ is the fading interference noise term and $\tilde{I}'_{i,k}$ is the term of a direct wave. With Eq. (\ref{eq:covariance}), the variances of them are expressed as
\begin{equation}
\begin{split}
\operatorname{Var}\{|\tilde{I}_{i,k}|\}=& \operatorname{E}_{\mathbf{b}_k,\tau_k}\left\{\int_{-\infty}^{\infty} \int_0^T \int_0^T  g_k(\tau) \right. \\
 \cdot & \xi_{i,k,k}(\tau+\tau_k, \tau+\tau_k; t_1,t_2)dt_1dt_2 d\tau \biggr\}\\
\operatorname{Var}\{|\tilde{I}'_{i,k}|\}=& \operatorname{E}_{\mathbf{b}_k,\tau_k}\left\{\int_0^T \int_0^T  \xi_{i,k,k}(\tau_k, \tau_k; t_1,t_2) dt_1dt_2\right\}.
\end{split}
\end{equation}
From the assumption 6 and 7, $\tau_k$ satisfies that $l_k T_c \leq \tau_k < (l_k + 1)T_c$, where $l_k$ $(0 \leq l_k < N)$ is an integer. The double integral term is written as
\begin{equation}
\begin{split}
&\int_0^T \int_0^T  \xi_{i,k,k}(\tau_k, \tau_k; t_1,t_2) dt_1dt_2\\
=& \left|R_{i,k}\left(\tau_k ,0, l_k\right) + \hat{R}_{i,k}\left(\tau_k, 0, l_k\right) \right|^2,
\end{split}
\end{equation}
where
 \begin{equation}
 \begin{split}
 R_{i,k}\left(\tau, n, l \right) =& \left(\tau - nT - lT_c\right)\\
\cdot& \left\{b_{k,-n-1} \sum_{m=1}^{l}\overline{s_{i,m}}s_{k,N-l+m}+ b_{k,-n}\sum_{m=1}^{N-l} \overline{s_{i,l+m}}s_{k,m} \right\},\\
 \hat{R}_{i,k}\left(\tau, n, l\right) =& \left(nT + (l+1)T_c - \tau\right)\\
\cdot& \left\{b_{k,-n-1} \sum_{m=1}^{l+1}\overline{s_{i,m}}s_{k,N-l+m-1}\right.\\
+& \left.b_{k,-n}\sum_{m=1}^{N-l-1} \overline{s_{i,l+m+1}}s_{k,m} \right\}.
 \end{split}
 \label{eq:cor2}
 \end{equation}
 Note that $R_{i,k}\left(\tau, n, l \right)$ and $\hat{R}_{i,k}\left(\tau, n, l\right)$ are crosscorrelation functions in a chip-synchronous CDMA model.
 We define
  \begin{equation}
 \Gamma_{i,k}(\tau,n,l) = \left|R_{i,k}\left(\tau ,n, l\right) + \hat{R}_{i,k}\left(\tau, n, l\right) \right|^2,
   \end{equation}
so that $\operatorname{Var}\{|\tilde{I}'_{i,k}|\}$ is concisely written as 
\begin{equation}
\begin{split}
&\operatorname{Var}\{|\tilde{I}'_{i,k}|\}\\
=& \frac{1}{T}\cdot \operatorname{E}_{\mathbf{b}_k}\left\{ \int_0^T\int_0^T \int_0^T  \xi_{i,k,k}(\tau_k, \tau_k; t_1,t_2) dt_1dt_2 d\tau_k \right\}\\
=&\frac{1}{T} \cdot \operatorname{E}_{\mathbf{b}_k}\left\{ \sum_{l_k=0}^{N-1}\int_{l_kT_c}^{(l_k+1)T_c}  \Gamma_{i,k}(\tau_k,0,l_k) d\tau_k \right\}.
\end{split}
\label{eq:tilde_I}
\end{equation}
 We consider the variance of $|\tilde{I}_{i,k}|$. Similar to the faded signal term, from the assumption 6, $\tau$ satisfies that $n'_k T + l'_k T_c \leq \tau < n'_kT + (l'_k + 1)T_c$, where $l'_k$ and $n'_k$ are the integers which satisfy $0 \leq l'_k < N$ and $n'_k \geq 0$. When we take the average over $\tau_k$, the double integral term in $\operatorname{Var}\{|\tilde{I}_{i,k}|\}$ is
  \begin{equation}
 \begin{split}
 &\frac{1}{T}\int^T_0\int_0^T \int_0^T \xi_{i,k,k}(\tau+\tau_k, \tau+\tau_k; t_1,t_2)dt_1dt_2 d\tau_k\\
  =&\frac{1}{T}\int_{\tau'}^{(l'_k+1)T_c} \Gamma_{i,k}\left(\tau_k, n'_k, l'_k\right)d\tau_k\\
  +&\frac{1}{T}\sum^{N-1}_{l=l'_k+1}\int_{lT_c}^{(l+1)T_c} \Gamma_{i,k}\left(\tau_k, n'_k, l\right)d\tau_k\\
  +&\frac{1}{T}\sum^{l'_k}_{l=0}\int_{lT_c}^{(l+1)T_c} \Gamma_{i,k}\left(\tau_k, n'_k+1, l\right) d\tau_k\\
 +&\frac{1}{T}\int_{l'_kT_c}^{\tau'}\Gamma_{i,k}\left(\tau_k, n'_k+1, l'_k\right)d\tau_k,
  \end{split}
  \label{eq:fad_i}
 \end{equation}
where $\tau' = \tau - n'_kT$.
From Eq. (\ref{eq:fad_i}), since each bit $b_{k,-n}$ is independent, it is sufficient to consider only the two adjacent bits in each term of Eq. (\ref{eq:fad_i}). In other words, it is sufficient to consider only the bits $b_{k,-n'_k}$ and $b_{k,-n'_k-1}$. Thus, we obtain
\begin{equation}
 \begin{split}
  &\frac{1}{T}\cdot\operatorname{E}_{\mathbf{b}_k}\left\{\int^T_0\int_0^T \int_0^T \xi_{i,k,k}(\tau+\tau_k, \tau+\tau_k; t_1,t_2)dt_1dt_2 d\tau_k \right\}\\
 =&\frac{1}{T} \cdot\operatorname{E}_{\mathbf{b}_k}\left\{\sum_{l_k=0}^{N-1}\int_{l_kT_c}^{(l_k+1)T_c} \Gamma_{i,k}\left(\tau_k, n'_k, l_k\right) d\tau_k \right\}\\
 =&\operatorname{Var}\{|\tilde{I}'_{i,k}|\}.
   \end{split}
\end{equation}
In the above equations, we have set $n'_k = 0$ to obtain the last equality.
Then, we can express $\operatorname{Var}\{|\tilde{I}_{i,k}|\}$ as the product of $\operatorname{Var}\{|\tilde{I}'_{i,k}|\}$ and the integral covariance term.
With the above results, we have
\begin{equation}
\operatorname{Var}\{|\tilde{I}_{i,k}|\} = \frac{1}{T}L_k \cdot\operatorname{E}_{\mathbf{b}_k}\left\{\sum_{l_k=0}^{N-1}\int_{l_kT_c}^{(l_k+1)T_c}\Gamma_{i,k} \left(\tau_k,0, l_k\right) d\tau_k \right\},
\end{equation}
where
\begin{equation}
L_k = \int_{-\infty}^{\infty}g_k(\tau)d \tau = \int_{0}^{M_kT}g_k(\tau)d \tau.
\end{equation}
In the worst case for $\operatorname{Var}\{F_i\}$, where $g_i(\tau)$ is the rectangular function, $L_k$ is
\begin{equation}
L_k = M_kC_kT.
\end{equation}
From these calculations, the variance of $I_i$ is
\begin{equation}
\begin{split}
\operatorname{Var}\{I_{i}\} =& \frac{P}{4T}\sum^K_{\substack{k=1 \\ k \neq i}} (1 + \gamma^2_kL_k)\\
& \cdot\operatorname{E}_{\mathbf{b}_k} \left\{ \sum_{l_k=0}^{N-1} \int_{l_kT_c}^{(l_k+1)T_c} \Gamma_{i,k}\left(\tau_k, 0, l_k\right)  d\tau_k \right\}.
\end{split}
\label{eq:inter_cor}
\end{equation}
To increase the lower bound of SNR, it is necessary to reduce the sum and integral term since $(1 + \gamma^2_kL_k)$ is constant.

\section{New Expression of SNR}
In this section, our goal is to calculate Eq. (\ref{eq:fad_cor}) and Eq. (\ref{eq:inter_cor}) and to derive the new expression of the upper bound of SNR. In \cite{basis}, it has been shown that the correlation of a chip-synchronous CDMA system can be written in a quadratic form. With this expression, Eq. (\ref{eq:cor}) is rewritten as
 \begin{equation}
 \begin{split}
 R_i\left(\tau, n, l \right) =& \left(\tau - nT - lT_c\right)\mathbf{s}^*_i B^{(l)}_{b_{i,-n-1},b_{i,-n}}\mathbf{s}_i,\\
 \hat{R}_i\left(\tau, n, l\right) =& \left(nT + (l+1)T_c - \tau\right)\mathbf{s}^*_i B^{(l+1)}_{b_{i,-n-1},b_{i,-n}}\mathbf{s}_i,\\
 \end{split}
 \label{eq:cor_quad}
 \end{equation}
 where $\mathbf{z}^*$ is a complex conjugate transpose of $\mathbf{z}$,
\begin{equation}
\mathbf{s}_k = (s_{k,1}, s_{k,2}, \ldots, s_{k,N})^\mathrm{T}
\end{equation}
and
\begin{equation}
B^{(l)}_{b_{k,n-1},b_{k,n}} = \left( \begin{array}{c c}
O & b_{k,n-1}E_{l} \\
b_{k,n}E_{N-l} & O
\end{array}
\right).
\end{equation}
In the above equations, $\mathbf{s}^\mathrm{T}$ is the transpose of $\mathbf{s}$ and $E_l$ is the identity matrix of size $l$. Similar to  Eq. (\ref{eq:cor_quad}), Eq. (\ref{eq:cor2}) is rewritten as
 \begin{equation}
 \begin{split}
 R_{i,k}\left(\tau, n, l \right) =& \left(\tau - nT - lT_c\right)\mathbf{s}^*_i B^{(l)}_{b_{k,-n-1},b_{k,-n}}\mathbf{s}_k,\\
 \hat{R}_{i,k}\left(\tau, n, l\right) =& \left(nT + (l+1)T_c - \tau\right)\mathbf{s}^*_i B^{(l+1)}_{b_{k,-n-1},b_{k,-n}}\mathbf{s}_k.\\
 \end{split}
 \label{eq:cor2_quad}
 \end{equation}
Thus, we can rewrite $\Gamma_{i,k}(\tau_k,0,l_k)$ in Eq. (\ref{eq:inter_cor}) as
 \begin{equation}
  \begin{split}
   &\Gamma_{i,k}(\tau_k,0,l_k) \\
=& \left| \left(\tau_k - l_kT_c\right) \mathbf{s}^*_i B^{(l_k)}_{b_{k,-1},b_{k,0}} \mathbf{s}_k \right. \left.+  \left((l_k+1)T_c - \tau_k\right)\mathbf{s}^*_i B^{(l_k+1)}_{b_{k,-1},b_{k,0}} \mathbf{s}_k \right|^2\\
 =& \left(\tau_k - l_kT_c\right)^2\left|\mathbf{s}^*_i B^{(l_k)}_{b_{k,-1},b_{k,0}} \mathbf{s}_k\right|^2 \\
 +& \left((l_k+1)T_c - \tau_k\right)^2\left|\mathbf{s}^*_i B^{(l_k+1)}_{b_{k,-1},b_{k,0}} \mathbf{s}_k \right|^2\\
 +&  2\left(\tau_k - l_kT_c\right) \left((l_k+1)T_c - \tau_k\right)\\
\cdot& \operatorname{Re}\left[\left(\mathbf{s}^*_i B^{(l_k)}_{b_{k,-1},b_{k,0}} \mathbf{s}_k\right) \overline{\left(\mathbf{s}^*_i B^{(l_k+1)}_{b_{k,-1},b_{k,0}} \mathbf{s}_k\right)} \right],
   \end{split}
 \end{equation}
Replacing $k$ with $i$, we obtain the expression of $\Gamma_i(\tau,0,l_i)$ in Eq. (\ref{eq:fad_cor}). 

It can be shown that $\mathbf{s}_k$ is expressed as \cite{basis}
\begin{equation}
\mathbf{s}_k = \frac{1}{\sqrt{N}}\sum_{m=1}^N \alpha^{(k)}_m\mathbf{w}_m(0)=\frac{1}{\sqrt{N}}\sum_{m=1}^N \beta^{(k)}_m\mathbf{w}_m\left(\frac{1}{2N}\right),
\end{equation}
where $\mathbf{w}_m(\eta)$ is the basis vector whose $n$-th component is expressed as
\[\left(\mathbf{w}_m(\eta)\right)_n = \exp\left(2\pi j (n-1)\left(\frac{m}{N} + \eta\right)\right),\]
$\alpha^{(k)}_m$ and $\beta^{(k)}_m$ are complex coefficients. 
There is the relations between $\alpha^{(k)}_m$ and $\beta^{(k)}_m$ such that
\begin{equation}
\begin{split}
{\boldsymbol \alpha^{(k)}} &= \Phi {\boldsymbol \beta^{(k)}},\\
 {\boldsymbol \beta^{(k)}} &= \hat{\Phi} {\boldsymbol \alpha^{(k)}},
\end{split}
\label{eq:relation_phi}
\end{equation}
where,
\begin{equation}
 {\boldsymbol \alpha^{(k)}} = \left( \begin{array}{c}
 \alpha_1^{(k)}\\
 \alpha_2^{(k)}\\
 \vdots\\
 \alpha_N^{(k)}
 \end{array} \right),  {\boldsymbol \beta^{(k)}} = \left( \begin{array}{c}
 \beta_1^{(k)}\\
 \beta_2^{(k)}\\
 \vdots\\
 \beta_N^{(k)}
 \end{array} \right),
  \end{equation}
  $\Phi$ and $\hat{\Phi}$ are the unitary matrices whose $(m,n)$-th components are
  \begin{equation}
  \begin{split}
   \Phi_{m,n}=&\frac{1}{N}\cdot\frac{2}{1-\exp(2 \pi j (\frac{n-m}{N} + \frac{1}{2N}))},\\
 \hat{\Phi}_{m,n}  =&\frac{1}{N}\cdot\frac{2}{1-\exp(2 \pi j (\frac{n-m}{N} - \frac{1}{2N}))}.
  \end{split}
  \end{equation}
Note that the vectors $\frac{1}{\sqrt{N}}\mathbf{w}_m(0)$ and $ \frac{1}{\sqrt{N}}\mathbf{w}_m(1/(2N))$ are eigenvectors of $B^{(l)}_{1,1}$ and $B^{(l)}_{-1,1}$ respectively. In other words, the matrices $B^{(l)}_{1,1}$ and $B^{(l)}_{-1,1}$ are decomposed as
\begin{equation}
\begin{split}
B^{(l)}_{1,1} & = V \Lambda^{(l)} V^*, \\
B^{(l)}_{-1,1} & = \hat{V} \hat{\Lambda}^{(l)} \hat{V}^*, 
\end{split}
\end{equation}
where
\begin{equation}
\begin{split}
V =& \frac{1}{\sqrt{N}} \left(\begin{array}{c c c c}
\mathbf{w}_1(0) & \mathbf{w}_2(0) & \cdots & \mathbf{w}_N(0)
\end{array}
\right),\\
\Lambda^{(l)} =& \operatorname{diag}\left(
\begin{array}{c c c c}
\lambda^{(l)}_1 & \lambda^{(l)}_2 & \cdots & \lambda^{(l)}_N
\end{array}\right),\\
\hat{V} =& \frac{1}{\sqrt{N}} \left(\begin{array}{c c c c}
\mathbf{w}_1\left(\frac{1}{2N}\right) & \mathbf{w}_2\left(\frac{1}{2N}\right) & \cdots & \mathbf{w}_N\left(\frac{1}{2N}\right)
\end{array}
\right),\\
\hat{\Lambda}^{(l)} =& \operatorname{diag}\left(
\begin{array}{c c c c}
\hat{\lambda}^{(l)}_1 & \hat{\lambda}^{(l)}_2 & \cdots & \hat{\lambda}^{(l)}_N
\end{array}\right).\\
\end{split}
\end{equation}
In the above equations, $\operatorname{diag}(\mathbf{z})$ is a diagonal matrix with the elements of vector $\mathbf{z}$ on the main diagonal and the eigenvalues $\lambda^{(l)}_m$ and $\hat{\lambda}^{(l)}_m$ are expressed as
\begin{equation}
\begin{split}
\lambda_m^{(l)} &=  \exp\left(-2 \pi j l\frac{m}{N}\right),\\
\hat{\lambda}_m^{(l)} &=  \exp\left(-2 \pi j l\left(\frac{m}{N} + \frac{1}{2N}\right)\right).
\end{split}
\end{equation}
Thus, the four types of the correlation $\mathbf{s}^*_i B^{(l_k)}_{b_{k,-1},b_{k,0}} \mathbf{s}_k$ are expressed as
\begin{equation}
\begin{split}
\mathbf{s}^*_i B^{(l)}_{1,1} \mathbf{s}_k &= \sum_{m=1}^N \lambda^{(l)}_m \overline{\alpha^{(i)}_m}\alpha^{(k)}_m,\\
\mathbf{s}^*_i B^{(l)}_{-1,-1} \mathbf{s}_k &= -\sum_{m=1}^N \lambda^{(l)}_m \overline{\alpha^{(i)}_m}\alpha^{(k)}_m,\\
\mathbf{s}^*_i B^{(l)}_{-1,1} \mathbf{s}_k &= \sum_{m=1}^N \hat{\lambda}^{(l)}_m \overline{\beta^{(i)}_m}\beta^{(k)}_m,\\
\mathbf{s}^*_i B^{(l)}_{1,-1} \mathbf{s}_k &= -\sum_{m=1}^N \hat{\lambda}^{(l)}_m \overline{\beta^{(i)}_m}\beta^{(k)}_m.
\end{split}
\label{eq:cor_decomp}
\end{equation}

\begin{figure}[htbp] %  figure placement: here, top, bottom, or page
   \centering
   \includegraphics[width=3in]{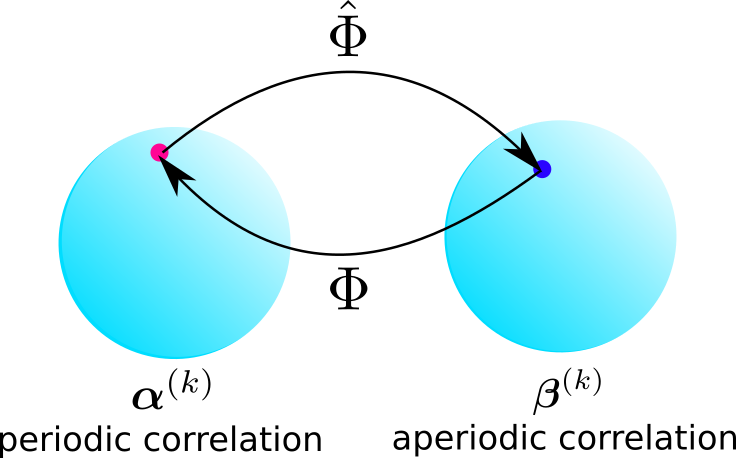} 
   \caption{The relation among the vectors of the coefficients on the hyperspheres}
   \label{fig:cor}
\end{figure}

From Eq. (\ref{eq:cor_decomp}), the coefficients $\alpha^{(k)}_m$ is related to the periodic correlation and $\beta^{(k)}_m$ is related to the aperiodic correlation. Since each of the vectors $\mathbf{w}_m(0)$ and $\mathbf{w}_m\left(\frac{1}{2N}\right)$ is orthogonal with respect to different $m$, we obtain the condition that
\begin{equation}
\left\|\boldsymbol \alpha^{(k)}\right\|^2 = \left\|\boldsymbol \beta^{(k)}\right\|^2=N.
\end{equation}
Therefore, the vectors of the coefficients, $\boldsymbol \alpha^{(k)}$ and $\boldsymbol  \beta^{(k)}$ locate on the hyperspheres. Figure \ref{fig:cor} shows the relation between the vectors of the coefficients, $\boldsymbol \alpha^{(k)}$ and $\boldsymbol \beta^{(k)}$. The Euclidian norm is preserved through the transformation since the matrices $\Phi$ and $\hat{\Phi}$. Therefore, the vector $\boldsymbol \alpha^{(k)}$ on a hypersphere is moved to $\boldsymbol \beta^{(k)}$ on another hypersphere by $\hat{\Phi}$. Similarly, the vector $\boldsymbol \beta^{(k)}$ on a hypersphere is moved to $\boldsymbol \alpha^{(k)}$ on another hypersphere by $\Phi$.

With these expressions, we calculate Eq. (\ref{eq:fad_cor}) and Eq. (\ref{eq:inter_cor}). First, Calculating the integral of $\Gamma_{i,k}(\tau_k,0,l_k)$, we have 
\begin{equation}
\begin{split}
&\int_{l_kT_c}^{(l_k+1)T_c} \Gamma_{i,k}(\tau_k,0,l_k) d\tau_k\\
 =& \frac{1}{3}T_c^3\left|\mathbf{s}^*_i B^{(l_k)}_{b_{k,-1},b_{k,0}} \mathbf{s}_k\right|^2 + \frac{1}{3}T_c^3\left|\mathbf{s}^*_i B^{(l_k+1)}_{b_{k,-1},b_{k,0}} \mathbf{s}_k \right|^2\\
 +& \frac{1}{3}T_c^3 \operatorname{Re}\left[\left(\mathbf{s}^*_i B^{(l_k)}_{b_{k,-1},b_{k,0}} \mathbf{s}_k\right) \overline{\left(\mathbf{s}^*_i B^{(l_k+1)}_{b_{k,-1},b_{k,0}} \mathbf{s}_k\right)} \right].
\end{split}
\label{eq:inter_term}
\end{equation}
When we take the average of Eq. (\ref{eq:inter_term}) over the bits $b_{k,-1}$ and $b_{k,0}$, the resultant averaged quantity is
\begin{equation}
\begin{split}
&\operatorname{E}_{\mathbf{b}_k}\left\{\left|\mathbf{s}^*_i B^{(l_k)}_{b_{k,-1},b_{k,0}} \mathbf{s}_k\right|^2\right\}\\
 =& \frac{1}{2}\left\{ \left| \sum_{m=1}^{N}\lambda_m^{(l_k)} \overline{\alpha^{(i)}_m}\alpha^{(k)}_m\right|^2 + \left| \sum_{m=1}^{N}\hat{\lambda}_m^{(l_k)} \overline{\beta^{(i)}_m}\beta^{(k)}_m\right|^2\right\},\\
&\operatorname{E}_{\mathbf{b}_k}\left\{\left|\mathbf{s}^*_i B^{(l_k+1)}_{b_{k,-1},b_{k,0}} \mathbf{s}_k\right|^2\right\}\\
 =& \frac{1}{2}\left\{ \left| \sum_{m=1}^{N}\lambda_m^{(l_k+1)} \overline{\alpha^{(i)}_m}\alpha^{(k)}_m\right|^2 + \left| \sum_{m=1}^{N}\hat{\lambda}_m^{(l_k+1)} \overline{\beta^{(i)}_m}\beta^{(k)}_m\right|^2\right\}\\
\end{split}
\end{equation}
and
\begin{equation}
\begin{split}
&\operatorname{E}_{\mathbf{b}_k}\left\{\operatorname{Re}\left[\left(\mathbf{s}^*_i B^{(l_k)}_{b_{k,-1},b_{k,0}} \mathbf{s}_k\right) \overline{\left(\mathbf{s}^*_i B^{(l_k+1)}_{b_{k,-1},b_{k,0}} \mathbf{s}_k\right)} \right]\right\}\\
=&\frac{1}{2} \operatorname{Re}\left[ \left(\sum_{m=1}^{N}\lambda_m^{(l_k)} \overline{\alpha^{(i)}_m}\alpha^{(k)}_m\right) \overline{\left(\sum_{m'=1}^{N}\lambda_m^{(l_k+1)} \overline{\alpha^{(i)}_{m'}}\alpha^{(k)}_{m'}\right)}\right] \\
+& \frac{1}{2} \operatorname{Re}\left[ \left(\sum_{m=1}^{N}\hat{\lambda}_m^{(l_k)} \overline{\beta^{(i)}_m}\beta^{(k)}_m\right) \overline{\left(\sum_{m'=1}^{N}\hat{\lambda}_m^{(l_k+1)} \overline{\beta^{(i)}_{m'}}\beta^{(k)}_{m'}\right)}\right].
\end{split}
\end{equation}
It is straightforward to show that
\begin{equation}
\begin{split}
\sum_{l_k=0}^{N-1}\lambda_m^{(l_k)}\overline{\lambda_{m'}^{(l_k)}} &= \sum_{l_k=0}^{N-1}\exp\left(2 \pi j l_k\frac{m'-m}{N}\right) = N\delta_{mm'},\\
\sum_{l_k=0}^{N-1}\hat{\lambda}_m^{(l_k)}\overline{\hat{\lambda}_{m'}^{(l_k)}} &= \sum_{l_k=0}^{N-1}\exp\left(2 \pi j l_k\frac{m'-m}{N}\right) = N\delta_{mm'},\\
\sum_{l_k=0}^{N-1}\lambda_m^{(l_k+1)}\overline{\lambda_{m'}^{(l_k+1)}} &= \sum_{l_k=0}^{N-1}\exp\left(2 \pi j \left(l_k+1\right)\frac{m'-m}{N}\right) = N\delta_{mm'},\\
\sum_{l_k=0}^{N-1}\hat{\lambda}_m^{(l_k+1)}\overline{\hat{\lambda}_{m'}^{(l_k+1)}} &= \sum_{l_k=0}^{N-1}\exp\left(2 \pi j \left(l_k+1\right)\frac{m'-m}{N}\right) = N\delta_{mm'},\\
\sum_{l_k=0}^{N-1}\lambda_m^{(l_k)}\overline{\lambda_{m'}^{(l_k+1)}} &= \exp\left(2 \pi j \frac{m'}{N}\right)\sum_{l_k=0}^{N-1}\exp\left(2 \pi j l_k\frac{m'-m}{N}\right)\\
& = N\exp\left(2 \pi j \frac{m'}{N}\right)\delta_{mm'},\\
\sum_{l_k=0}^{N-1}\hat{\lambda}_m^{(l_k)}\overline{\hat{\lambda}_{m'}^{(l_k+1)}} &= \exp\left(2 \pi j \left(\frac{m'}{N} + \frac{1}{2N}\right)\right)\\
&\cdot\sum_{l_k=0}^{N-1}\exp\left(2 \pi j l_k\frac{m'-m}{N}\right)\\
& = N \exp\left(2 \pi j \left(\frac{m'}{N} + \frac{1}{2N}\right)\right)\delta_{mm'},
\end{split}
\label{eq:thm_lk}
\end{equation}
where
\[\delta_{mm'} = \left\{ \begin{array}{c c}
1 & m=m'\\
0 & m \neq m'
\end{array}
\right. .
\]
From Eq. (\ref{eq:thm_lk}), we can calculate the sum and integral term of Eq. (\ref{eq:inter_cor}). Then, we rewrite Eq. (\ref{eq:inter_cor}) as
\begin{equation}
\operatorname{var}\{I_{i}\} = \frac{PT^2}{12N^2}\sum_{\substack{k=1 \\ k \neq i}}^K (1 + \gamma_k^2 L_k)\sum_{m=1}^N S^{i,k}_m,
\label{eq:inter_new}
\end{equation}
where
\begin{equation}
\begin{split}
S^{i,k}_m &= \left|\alpha_m^{(i)}\right|^2\left|\alpha_m^{(k)}\right|^2\left(1 + \frac{1}{2}\cos\left(2 \pi \frac{m}{N}\right)\right)\\
&+ \left|\beta_m^{(i)}\right|^2\left|\beta_m^{(k)}\right|^2\left(1 + \frac{1}{2}\cos\left(2 \pi \left(\frac{m}{N} + \frac{1}{2N}\right)\right)\right).
\end{split}
\label{eq:S}
\end{equation}
Replacing $k$ with $i$, we obtain the expression of $\Gamma_i(\tau,0,l_i)$, which is defined in Eq. (\ref{eq:fad_cor}). Then, Eq. (\ref{eq:fad_cor}) is rewritten as
\begin{equation}
\operatorname{Var}\{F_i\} \leq \frac{PT^3}{12N^2}\gamma_i^2 C_i M_i \sum_{m=1}^N S_m^{i,i}.
\label{eq:fad_new}
\end{equation}
From the above expressions, we arrive at the formula for the lower bound of SNR of the user $i$
\begin{equation}
\operatorname{SNR}_i \geq \left\{\frac{1}{6N^2}\sum_{k=1}^K Z_{i,k} \sum_{m=1}^M S_m^{i,k} + \frac{N_0}{2PT}\right\}^{-1/2},
\label{eq:SNR}
\end{equation}
where
\[ Z_{i,k} = \left\{ \begin{array}{c c}
\displaystyle \gamma_i^2 C_i M_iT & k=i\\
\displaystyle 1 + \gamma_k^2 L_k & k \neq i
\end{array}
\right. .\]
When $Z_{i,k} = 1$ $(i \neq k)$ and  $Z_{i,i} = 0$, Eq. (\ref{eq:SNR}) is equivalent to the expression in \cite{mazzini} and \cite{pursley}. 

Equation (\ref{eq:S}) shows that there is the relation among SNR and periodic and aperiodic correlation. The first term of Eq. (\ref{eq:S}) is related to the periodic correlation since the coefficients $\alpha_m^{(k)}$ appear in the periodic correlation. The last term of Eq. (\ref{eq:S}) is related to the aperiodic correlation since the coefficients $\beta_m^{(k)}$ appear in the aperiodic correlation. In Section V, we discuss the relation between SNR and correlation in detail.

The parameters of the spreading sequences $\alpha^{(k)}_m$ and $\beta^{(k)}_m$ are complex. Thus, the coefficients $\alpha^{(k)}_m$ and $\beta^{(k)}_m$ are divided into the real parts and imaginary parts. We define
\begin{equation}
\begin{split}
\alpha^{(k)}_{1,m} &= \operatorname{Re}[\alpha^{(k)}_m],\\
\alpha^{(k)}_{2,m} &= \operatorname{Im}[\alpha^{(k)}_m],\\
\beta^{(k)}_{1,m} &= \operatorname{Re}[\beta^{(k)}_m],\\
\beta^{(k)}_{1,m} &= \operatorname{Im}[\beta^{(k)}_m].
\end{split}
\label{eq:re_im}
\end{equation}
From Eq. (\ref{eq:re_im}), Eq. (\ref{eq:S}) is rewritten as
\begin{equation}
\begin{split}
\hat{S}_m^{i,k}&= \left(\left(\alpha^{(i)}_{1,m}\right)^2 +\left(\alpha^{(i)}_{2,m}\right)^2  \right) \left(\left(\alpha^{(k)}_{1,m}\right)^2 +\left(\alpha^{(k)}_{2,m}\right)^2  \right)\\
&\cdot \left(1 + \frac{1}{2}\cos\left(2\pi\frac{m}{N}\right)\right)\\
&+\left(\left(\beta^{(i)}_{1,m}\right)^2 +\left(\beta^{(i)}_{2,m}\right)^2  \right)\left(\left(\beta^{(k)}_{1,m}\right)^2 +\left(\beta^{(k)}_{2,m}\right)^2  \right)\\
&\cdot \left(1 + \frac{1}{2}\cos\left(2\pi\left(\frac{m}{N} + \frac{1}{2N}\right)\right)\right).
 \end{split}
 \label{eq:S2}
\end{equation}
Note that Eq. (\ref{eq:S2}) is differentiable with the parameters of the spreading sequence, $\alpha^{(k)}_{1,m}$, $\alpha^{(k)}_{2,m}$, $\beta^{(k)}_{1,m}$ and $\beta^{(k)}_{2,m}$. Thus, we can maximize SNR of the user $i$ with the Lagrange multipliers method. Although Eq. (\ref{eq:S2}) is convex in only the parameters of the user $k$, $\alpha^{(k)}_{1,m}$, $\alpha^{(k)}_{2,m}$, $\beta^{(k)}_{1,m}$ and $\beta^{(k)}_{2,m}$, it is not convex in all the parameters. In Part II, we discuss how we to the optimization problem.

\section{Relation between SNR and Mean-Square Correlation}
We show the relation between the expression of SNR and the mean-square correlations. The mean-square correlations are proposed as indices for performance of spreading sequences. In particular, the mean-square crosscorrelation is used for advantage of spreading sequences \cite{meansquare}. The mean-square crosscorrelation function $R_{CC}$ and the mean-square autocorrelation function $R_{AC}$ are defined as \cite{fzcfam} 
\begin{eqnarray}
R_{CC} &=& \frac{1}{K}\sum_{i=1}^K R^{(i)}_{CC} \label{eq:RCC0} \\
R_{AC} &=& \frac{1}{K}\sum_{i=1}^K R^{(i)}_{AC},\label{eq:RAC0} 
\end{eqnarray}
where
\begin{eqnarray}
R^{(i)}_{CC} &=& \frac{1}{(K-1)}\frac{1}{N^2} \sum_{\substack{k=1 \\ k \neq i}}^K \sum_{l=1-N}^{N-1}|C_{i,k}(l)|^2 \label{eq:RCC} \\
R^{(i)}_{AC} &=& \frac{1}{N^2} \sum_{l=1-N, l \neq 0}^{N-1}|C_{i}(l)|^2,\label{eq:RAC} 
\end{eqnarray}
\begin{equation}
C_{i,k}(l) = \left\{ \begin{array}{c c}
\displaystyle \sum_{n=1}^{N-l}\overline{s_{i,n+l}}s_{k,n} & 0 \leq l \leq N-1\\
\displaystyle \sum_{n=1}^{N+l}\overline{s_{i,n}}s_{k,n-l} & 1-N \leq l < 0\\
0 & \mbox{otherwise}
\end{array}\right.  
\end{equation}
and $C_{i}(l)$ is denoted by $C_{i,i}(l)$.
From \cite{meansquare}, Eqs. (\ref{eq:RCC}) and (\ref{eq:RAC}) is rewritten as
\begin{eqnarray}
R^{(i)}_{CC} &=& \frac{1}{2(K-1)}\frac{1}{N^2}  \sum_{\substack{k=1 \\ k \neq i}}^K \sum_{l=0}^{N-1}\left\{ |\theta_{i,k}(l)|^2 +|\hat{\theta}_{i,k}(l)|^2\right\},  \label{eq:RCC2} \\
R^{(i)}_{AC} &=& \frac{1}{2N^2}  \sum_{l=1}^{N-1}\left\{ |\theta_{i}(l)|^2 +|\hat{\theta}_{i}(l)|^2\right\}, \label{eq:RAC2} 
\end{eqnarray}
where $\theta_{i,k}(l)$ and $\hat{\theta}_{i,k}(l)$ are periodic and aperiodic correlation functions which are defined as
\begin{eqnarray}
\theta_{i,k}(l) &=& C_{i,k}(l) + C_{i,k}(l-N), \\
\hat{\theta}_{i,k}(l)&=& C_{i,k}(l) - C_{i,k}(l-N),
\end{eqnarray}
and $\theta_{i}(l)$ and $\hat{\theta}_{i}(l)$ are denoted by $\theta_{i,i}(l)$ and $\hat{\theta}_{i,i}(l)$. From $\cite{basis}$, $\theta_{i,k}(l)$ and $\hat{\theta}_{i,k}(l)$ are expressed as
\begin{eqnarray}
\theta_{i,k}(l) &=& \sum_{m=1}^N \lambda_m^{(l)}\overline{\alpha^{(i)}_m}\alpha_m^{(k)}, \\
\hat{\theta}_{i,k}(l)&=& \sum_{m=1}^N \hat{\lambda}_m^{(l)}\overline{\beta^{(i)}_m}\beta_m^{(k)}.
\end{eqnarray}
With these expressions, Eqs. (\ref{eq:RCC2}) and (\ref{eq:RAC2}) are rewritten as
\begin{equation}
\begin{split}
R^{(i)}_{CC} =& \frac{1}{2(K-1)}\frac{1}{N}\\
& \cdot \sum_{\substack{k=1 \\ k \neq i}}^K \sum_{m=1}^{N}\left\{ |\alpha^{(i)}_m|^2|\alpha^{(k)}_m|^2 + |\beta^{(i)}_m|^2|\beta^{(k)}_m|^2  \right\},
\label{eq:RCC3} 
\end{split}
\end{equation}
\begin{equation}
R^{(i)}_{AC} = \frac{1}{2N} \sum_{m=1}^{N}\left\{ |\alpha^{(i)}_m|^4  + |\beta^{(i)}_m|^4 \right\} - 1. 
\label{eq:RAC3} 
\end{equation}
In obtaining the above equations, we have used the relation that $|\theta_{i,k}(0)|^2 = |\hat{\theta}_{i,k}(0)|^2 = N^2$. Thus, we derive the bounds of Eq. (\ref{eq:SNR}) as
\begin{equation}
\begin{split}
& \left[\frac{1}{2N}\left\{Z_{i,i}(R^{(i)}_{AC} + 1) + Z_{i,U}(K-1)R_{CC}^{(i)} \right\} + \frac{N_0}{2PT}\right]^{-1/2}\\
\leq &\left\{\frac{1}{6N^2}\sum_{k=1}^K Z_{i,k} \sum_{m=1}^M S_m^{i,k} + \frac{N_0}{2PT}\right\}^{-1/2}\\
\leq &\left[\frac{1}{6N}\left\{Z_{i,i}(R^{(i)}_{AC} + 1) + Z_{i,L}(K-1)R_{CC}^{(i)} \right\} + \frac{N_0}{2PT}\right]^{-1/2},
\end{split}
\label{eq:rac_rcc}
\end{equation}
where $\displaystyle Z_{i,U} = \max_{k} Z_{i,k}$ and $\displaystyle Z_{i,L} = \min_{k} Z_{i,k}$. In the above inequality, we have used $-1 \leq \cos(x) \leq 1$. Note that the term of $R^{(i)}_{AC}$ corresponds to the faded signal term obtained in Eq. (\ref{eq:fad_new}). Moreover, the term of $R^{(i)}_{CC}$ corresponds to the interference noise term obtained Eq. (\ref{eq:inter_new}). Therefore, Eq. (\ref{eq:rac_rcc}) shows that the effects of faded signals and interference noise are reduced when the mean-square correlations $R^{(i)}_{AC}$ and $R^{(i)}_{CC}$ are small, respectively. Thus, it is necessary for designing spreading sequences to reduce mean-square correlations, $R^{(i)}_{AC}$ and $R^{(i)}_{CC}$.

\section{Conclusion}
We have shown the new expression of SNR whose terms are explicitly related to periodic and aperiodic correlation. This expression has been obtained from the quadratic forms of the correlation. With this expression, we have evaluated the lower bound of SNR. In Section V, we have shown the relation between SNR and mean-square correlations. This result shows that SNR will become higher when we resuce mean-square auto correlation and crosscorrelation low. 

In Part II, we construct the optimization problem: minimize the lower bound of SNR. Then, we derive the necessary conditions for the global solution and evaluate . 

A remaining issue is to obtain the better expression of SNR. We have used the relation that $g_i(\tau)$ has the upper bound. From this reason, our upper bound is rough. Our expression will get better if we have knowledge of the form of $g_i(\tau)$. Then, our lower bound of SNR will approach to equality.

\appendices
\section{}
In this appendix, we show how to obtain the condition that
\begin{equation*}
\sum_{n=1}^N \left|s_{k,n}\right|^2 = N.
\end{equation*}
We assume that the power of the transmitted signal is finite, and denote it by $P$. Thus, we obtain the follow equation
\begin{equation}
\begin{split}
&\frac{1}{T}\int^{T}_0|\zeta_k(t)|^2 dt\\
 =& \frac{P}{T}\int^{T}_0|s_k(t)|^2|b_k(t)|^2dt \\
+& \frac{P}{T}\int^{T}_0 \operatorname{Re}[\left(s_k(t)b_k(t)\right)^2\exp(2j \omega_c t + 2j\theta_k)]dt\\
=&\frac{P}{T}T_c\sum_{n=1}^N \left|s_{k,n}\right|^2\\
+& \frac{P}{T}\int^{T}_0 \operatorname{Re}[\left(s_k(t)\right)^2\exp(2j \omega_c t + 2j\theta_k)]dt\\
=&P.
\end{split}
\label{eq:power_s}
\end{equation}
In the above equation,  we have used Eq. (\ref{eq:thm1}) and the relation that $b_{k,n}^2=1$.

With a low pass filter, we can ignore the double frequency term \cite{pursley}, that is,
\begin{equation}
\int^{T}_0 \operatorname{Re}[\left(s_k(t)\right)^2\exp(2j \omega_c t + 2j\theta_k)]dt = 0.
\end{equation}
From Eq. (\ref{eq:power_s}), we have that
\begin{equation}
\sum_{n=1}^N \left|s_{k,n}\right|^2 = N.
\end{equation}
This is equivalent to
\begin{equation}
\sum_{n=1}^N \left|s_{k,n}\right|^2 = \left\|\mathbf{s}_k\right\|^2 = \left\|\boldsymbol \alpha^{(k)}\right\|^2 = \left\|\boldsymbol \beta^{(k)}\right\|^2=N,
\end{equation}
where $\| \cdot \|$ is the Euclidian norm. 

\section{}
In this appendix, we prove that the covariance function $g_k(\tau)$ has an upper bound. We define the inverse Fourier transformation of $h_k(\tau,t)$ as
\begin{equation}
h_k(f,t) = \int_{-\infty}^{\infty} H_k(f,t) \exp(2 \pi j f \tau) df.
\label{eq:fourier}
\end{equation}
From the assumption 1, the function $h_k(\tau,t)$ is transformed to $H_k(f,t)$ by the Fourier transformation and $H_k(f,t)$ is transformed to $h_k(\tau,t)$ by the inverse Fourier transformation, that is, 
\begin{equation}
\int_{-\infty}^{\infty}|h_k(\tau,t)|d\tau < \infty,
\end{equation}
\begin{equation}
\int_{-\infty}^{\infty}|H_k(f,t)|df < \infty,
\end{equation}
for all $t$. Then, the absolute value of the function $h_k(\tau,t)$ has a upper bound
\begin{equation}
\begin{split}
|h_k(\tau,t)| &= \left|\int_{-\infty}^{\infty} H_k(f,t) \exp(2 \pi j f \tau) df \right|\\
&\leq \int_{-\infty}^{\infty} |H_k(f,t)|df < \infty
\end{split}
\end{equation}
for all $\tau$ and $t$. Thus, the covariance function $g_k(t)$ is evaluated as
\begin{equation}
\begin{split}
g_k(\tau) &= \operatorname{E}\{h_k(\tau,0)\overline{h_k(\tau,0)}\}\\
&= \operatorname{E}\{|h_k(\tau,0)|^2\}\\
&\leq \operatorname{E}\{C\}\\
&= C< \infty,
\end{split}
\end{equation}
where
\begin{equation}
C = \sup_{\tau}|h_k(\tau,0)|^2 < \infty.
\end{equation}
In the above equations, we have used the assumption that $g_k(\tau)$ is independent of the variable $t$, that is, the channel is a selective fading channel.
We thus proved that $g_k(\tau)$ has an upper bound.

\section*{Acknowledgment}
The one of the authors, Hirofumi Tsuda, would like to thank for advice of Dr. Shin-itiro Goto.
\end{document}